# Language Interoperability in Control Network Programming

Kostadin Kratchanov[1], Efe Ergün[2]
[1,2]Yaşar University, Izmir, Turkey
([1]kostadin@kratchanov.net, [2]efeergun96@gmail.com)

***Abstract*** - Control Network Programming (CNP) is a programming paradigm which is being described with the maxim "Primitives + Control Network = Control Network program". It is a type of graphic programming. The Control Network is a recursive system of graphs; it can be a purely descriptive specification of the problem being solved. Clearly, 'drawing' the control network does not include any programming. The Primitives are elementary, easily understandable and clearly specified actions. Ultimately, they have to be programmed. Historically, they are usually coded in Free Pascal. The actual code of the primitives has never been considered important. The essence of an "algorithm" is represented by its control network. CNP was always meant to be an easy and fast approach for software application development that actually involves very little real programming.

Language interoperability (using different languages in the same software project) is a distinguished current trend in software development. It is even more important and natural in the case of CNP than for other programming paradigms. Here, interoperability practically means the possibility to use primitives written in various programming languages. The current report describes our first steps in creating applications using a multi-language set of primitives. Most popular and interesting programming languages have been addressed: Python, Java, and C. We show how to create applications with primitives written in those 'non-native' languages. We consider examples where the primitives in all those four programming languages are simultaneously used (multiple-language CNP). We also discuss CNP programming without programming (language-free CNP).

## I. Language interoperability in modern programming practice

Programming language interoperability is the ability of codes written in two or more programming languages to interact as part of the same system. Frequently, this involves passing messages and data between potentially very different languages and poses substantial problems. The concept has attracted increasing attention in the recent years (e.g., [1-5]). Its importance is widely recognized and accepted.

There is a number of reasons why language interoperability is highly desirable. For example, if a programmer has to implement a specific feature that has been already implemented in another language the corresponding program component can simply be reused. Some languages are especially fit and effective in implementing specific features and often have emerged to target particular problem domains. Rich third-party packages for certain languages are available. Also, every programmer usually has a preferred language in which their expertise and efficiency is better. There are hundreds of programming languages used and constantly being developed thus making language interoperability a necessity. Programmers with experience and preferences in different programming languages can easier team up for solving complex tasks.

Several tools and approaches have emerged to address different aspects of cross-language communication. Prominent examples are virtual machines and most notably the Java Virtual Machine (JVM) and .NET's Common Language Runtime (CLR), and markup languages notably the Extensible Markup Language (XML) and Starlink. Cross-platform integration between the virtual machine platforms has also been addressed. A virtual machine suggests the usage of an intermediate language such that a programmer can build a component using the language of their choice and this component will be compiled into the intermediate language; then an application written possibly in a different language can use this component without knowing what language was originally used to create the component. While the above approaches aim at integrating compiled languages, special challenges presents the interoperability with interpreted languages.

An important aspect of language interoperability is the seamless exchange of data between the components which means ensuring that the type systems of the corresponding languages are respected and that information is not lost when data moves between statically-typed and dynamically-typed languages. Typical methods for achieving this is the usage of metadata and standardized type systems.

As W. Toll [6] notices, "… developers are frequently looking for integrated developer environments (IDEs) that … offer a broad range of features and supported languages…. Some IDEs support practically every language known to man either natively or by extending capabilities with plug-ins or add-ons". Phrases like multi-language platforms and multi-language IDEs are being used.

## II. CONTROL NETWORK PROGRAMMING

Control Network Programming, or CNP means "programming through control networks" [7]. The following subsection is a short introduction to the main ideas and features of this novel programming approach.

### A. Introduction to CNP

The fundamental part of a CNP application is the Control Network (CN). It is a system of graphs called subnets. The subnets can call each other. One of them is identified as the main subnet. Each subnet consists of nodes (states) and arrows. A subnet has a unique initial state, and a number of final nodes. The arrows of a subnet are labeled with sequences of simple actions called primitives. Primitives are defined in the second major part of the application. Thus, a general maxim has been deducted: "Primitives + Control Network = Control Network Programming".

The CN may be of nondeterministic nature; an interpreter (inference engine) must implement a strategy for search/inference/computation in the CN. The system attempts to find a path from the initial node of the main subnet to a final node, executing the primitives along the way. This process may involve invoking other subnets. The execution of a primitive might result in failure in which case the system starts executing the already passed primitives of the arrow backwards. Upon the return to the source node of the arrow another outgoing arrow is attempted. In case no more un-attempted arrows exist the control backtracks; the corresponding actions are revoked. No interpreter exists in reality, instead, the CN is "compiled" into an intermediate program which embodies the CN together with the search process on it.

More details about CNP and the technicalities of the "execution" of a CN program can be found, e.g., in [7-11].

For better clarifying the further exposition, it would be helpful to present here the general structure of a CNP application. It is shown in Fig. 1. The CN is drawn and edited in the graph editor of the CNP IDE *SpiderCNP*, with the final result of this process being the file *SpiderNet.txt* which specifies the CN. The primitives (and the data structures) are defined in the file *SpiderUnit.pas*. As emphasized earlier, CNP = Primitives + CN, and the two files mentioned are the ones that the CNP programmer produces.

### B. Monkey and Banana – an exemplary CNP application

The Monkey and Banana problem is a famous toy problem in AI. As an exemplary CNP application, we show below a possible CNP solution of a simple version of this problem. Later, we will apply interoperability to this exemplary problem. The screenshots are from the *SpiderCNP* IDE.

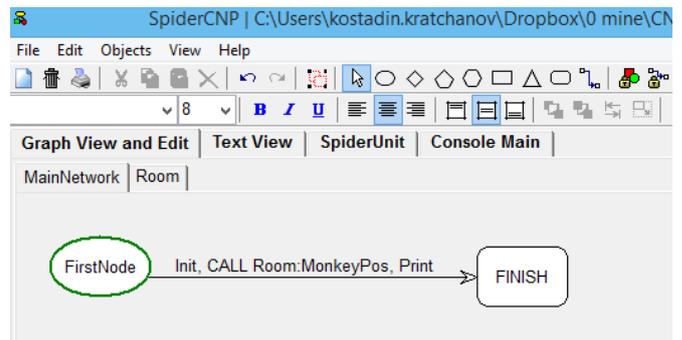

Figure 2. CN – main subnet

The presented CNP solution is purely declarative. The CN consists of two subnets shown in Fig. 2 and Fig. 3. Subnet *Room* of the CN corresponds to the plan of the room. Three positions in the room are identified: the door, a window, and the middle of the room. It is assumed that the banana hangs from the ceiling in the middle of the room.

Two possible executions are shown in Fig. 4. The system asks where the initial position of the monkey is; the answer is that the monkey is at the door. Then the system asks where initially the box is, and the answer is "Window" or "Door", respectively. Next the system finds two possible solutions to the problem for the specified initial positions of the monkey and the box in the second case, and one solution in the first.

Five primitives are used: *Init, Push, Walk, Climb,* and *Print*. They are defined in the file *SpiderUnit.pas*. The primitives are coded in Pascal. Their code is shown in Fig. 5 and Fig. 6.

The Monkey and Banana example above has been implemented in a 'classical' CNP manner. The implementation is declarative – the *Room* subnet is, in essence, a plan of the room with specification of the actions that the monkey can perform moving from one particular point in the room to another. The search for a solution is completely left for the system.

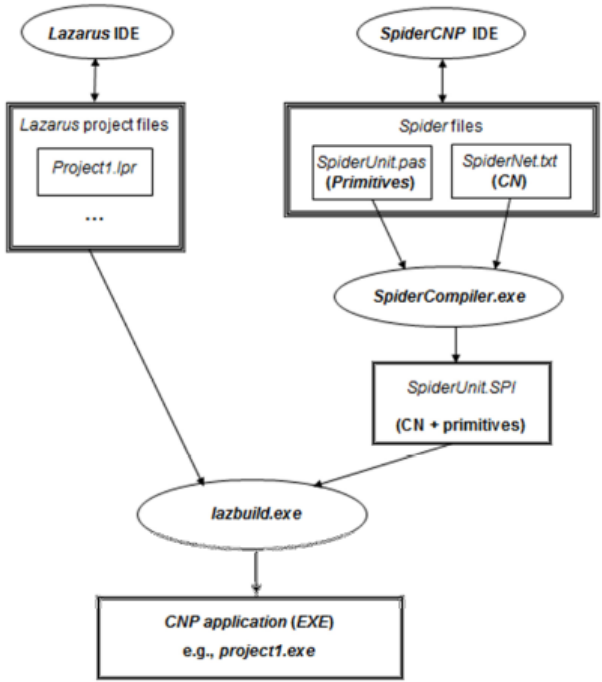

Figure 1. Structure of a CNP application

In the example, all five primitives used are written in Pascal. As a demonstration of interoperability in CNP, a modified version of this example is presented in Section V.B where the primitives are written in four different programming languages: one primitive in Pascal, Python and Java each, and two primitives in C.

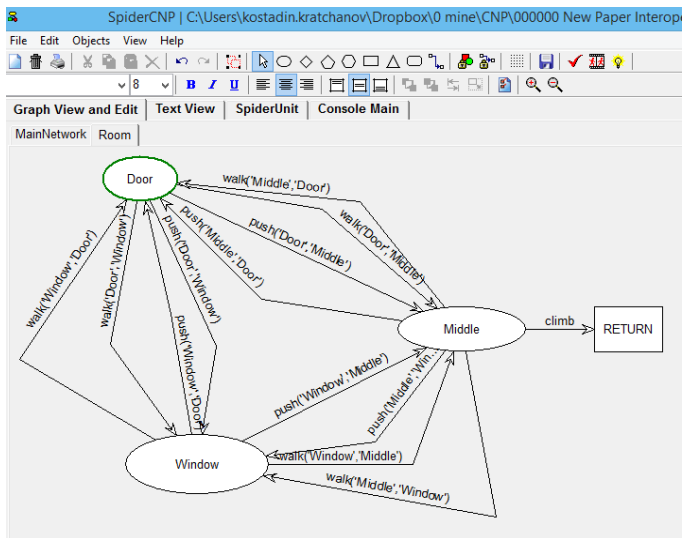

Figure 3.  CN – Subnet Room

Figure 4.  Two possible executions

## III. LANGUAGE INTEROPERABILITY IN CNP

CNP has been developed as an approach to programming that actually involves almost no coding. Language interoperability is intrinsically natural to it. We discuss below this nature of CNP and describe how in practice interoperability can be implemented in it.

### A. CNP is fast application development with almost no programming

CNP is a programming paradigm that integrates and extends declarative programming, imperative programming, and programming rule-based systems (e.g., [7]). Although universal, it is especially effective when solving problems which can be naturally represented in a graph-like manner, and/or whose descriptions exhibit nondeterminism or randomness. It is a genuine graphical programming.

As already discussed, its essence can be summarized by the maxim "Primitives + Control Network = Control Network program". The Control Network (CN) is a recursive system of graphs; it can be a purely descriptive specification of the problem. Clearly, it does not include any programming. The Primitives are the imperative part of the CN program.

CNP was created, and has always been applied, with the vision that the primitives are elementary, easily understandable and clearly specified actions. Ultimately, they have to be programmed (coded). Historically, they are usually coded in the Free Pascal language used in the Lazarus IDE [14-17], or in Delphi Pascal. However, the actual code of the primitives has never been considered important. The essence of an algorithm" is represented by its CN. This is a main concept underlying CNP. A CN program should be "natural", easily understandable, easily verifiable and easily testable by the user. We actually don't often use the phrase CNP programmer but rather CNP user. There is almost no programming (coding) in the usual sense involved in the process of developing a CNP application.

Generally, the development process is for the user to describe the problem through the CN using some simple, elementary primitives. The main stated advantage of CNP is that the development process is easy, fast, and easily verifiable. Basically, the CNP is not considered to be a "real" programming. It is more like "assembling", describing a CN. Even if some primitives must be coded this is definitely not a heavy programming.

In particular, if the problem in hand is nondeterministic, the user does not have to care about organizing the search. The "execution" of the CN is in fact itself a built-in local search process. Nondeterministic algorithms are directly "programmed" [12]. Numerous "declarative" tools for controlling and directing the built-in search are available to the user.

From the very beginning [12], as a new programming paradigm, CNP was meant to be an easy and fast approach for software application development that actually involves very little real programming.

Figure 5. Primitives I

Figure 6. Primitives II

*B. Programming without programming in CNP (language-free CNP)*

It was mentioned as early as in [8, 12] that a CNP user can use well documented libraries of simple primitives typical for the corresponding problem domain. In such a case, CNP will actually involve no coding at all. Typically, CNP applications use a very small number of primitives. Various references to some CNP applications are listed in [7].

We can call this approach CNP without programming, or language-free CNP.

We would like to remind here the ideas from the famous early resource [13].

*C. Interoperability in CNP (multiple-language CNP)*

Interoperability is even more important and natural in the case of CNP than for other programming paradigms.

Well, sometimes a user will still need to write their own primitives or extend the used libraries of primitives. Also, for some tasks a particular programming language may be more appropriate and efficient. Or simply, a user knows or prefers a language different from Pascal. All the consideration regarding language interoperability from Section I will be generally applicable.

As discussed in [7], the most powerful currently IDE for CNP is *SpiderCNP*, with versions for Delphi and Lazarus. The corresponding programming languages in which they are written are Delphi Pascal and Free Pascal (Object Pascal). Therefore, Pascal is the native language for writing primitives. Unfortunately, popularity of Pascal has decreased and currently Pascal only takes $10^{th}$ - $12^{th}$ position in popularity. Our studies among students using CNP regarding the difficulties they meet also show that giving the users the opportunity to write primitives using different programming languages is of high priority in order to extend and simplify the usage of CNP.

As mentioned already, a user developing CNP applications needs programming almost exclusively only when coding new primitives.

Interoperability in CNP practically means the possibility to use primitives written in different programming languages! The current report describes our first steps in creating applications using a multi-language set of primitives. The most popular and interesting programming languages have been addressed: Python, Java, C/C++.

The depth and sophistication of the interoperability in CNP presented below is far from the one achieved over years in the highly developed today general area of programming interoperability reviewed in Section I. In particular this concerns the issue of interchanging data between primitives of different languages and between the primitives and the main system written in Pascal. Future research in this area is needed.

## IV. USING SPECIFIC LANGUAGES IN CNP

The specifics of using some most popular [18] programming languages (C, Python, and Java) to write primitives in CNP are discussed below. As an example, a CNP application is used which calculates the well-known Body Mass Index (BMI) [19]. Two primitives are used: *BmiW* and *BmiWO*. *BmiW* has two integer parameters – the weight and the height of a person; it calculates and prints the corresponding BMI. *BmiWO* has a similar functionality but has no parameters - the input data is provided by the user. Actually, three projects have been created – one for each of the three languages considered. Both primitives in the corresponding CNP solution are written in the same language. The only purpose of the example chosen is to demonstrate the usage of non-native programming languages in CNP.

It should be emphasized that in all the three projects, regardless of the programming language used for writing the primitives, the CN remains the same. This CN is shown in Fig. 7. The outputs produced are also identical. Fig. 8 shows an examplary output for these CNP applications.

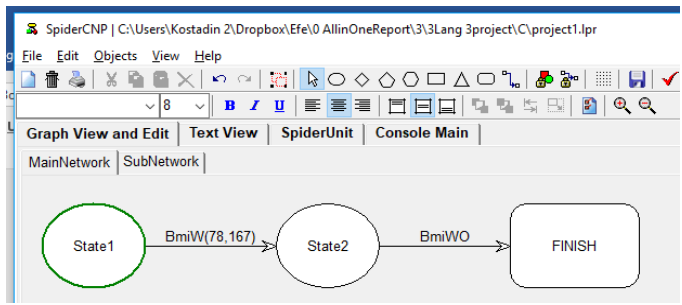

Figure 7.   BMI – CN

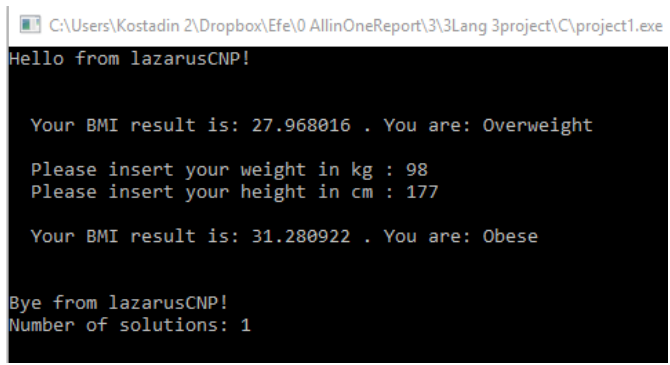

Figure 8.   BMI – Output

### A. C in CNP

The procedure for using primitives based on C functions is the following. A C file, *BMI.c* without a main section should be written, as illustrated below:

```c
#include <stdio.h>
#include <stdlib.h>
void BmiWc(int kg, int cm){
    double m = cm;
    m = m / 100;
    double bmi = ( kg / (m * m) );
    printf("\n");
if(bmi < 19){
    printf("  Your BMI result is: %f. You are: Thin ",bmi);
}else if(bmi < 25){
    printf("  Your BMI result is: %f. You are: Healthy ",bmi);
}else if(bmi < 30){
    printf("  Your BMI result is: %f. You are: Overweight ",bmi);
}else{
    printf("  Your BMI result is: %f. You are: Obese ",bmi);
}
printf(" \n ");
}
void BmiWOc(){
int cm,kg;
printf("\n");
printf("  Please insert your weight in kg : ");
scanf("%d" , &kg);
printf("  Please insert your height in cm : ");
scanf("%d" , &cm);
BmiWc(kg,cm);
}
```

The C file is then compiled. For example, one who has a GNU compiler on their PC can open the Windows PowerShell at the same location where the C file is saved, and type "*gcc -c BMI.c*". This will produce a file *BMI.o* which we must import into the CNP primitive.

We now need to open our CNP project and create a new unit, called here *Unit1* which wraps the created .*o* file into a Pascal unit [20] and looks as shown below. Note that the library *libmsvcrt* is required, as well as the unit *CTypes*.

```
unit Unit1;
{$mode objfpc}{$H+}
{$link Bmi.o}
{$linklib libmsvcrt}
interface
uses
  Classes, SysUtils, CTypes;
  procedure BmiWO; cdecl; external;
  procedure BmiW(kg : CTypes.cint32;m : CTypes.cint32); cdecl; external;
implementation
end.
```

As a result, Pascal procedures *BmiWO* and *BmiW* are defined which use the compiled already C functions *BmiWO*

and *BmiW*. No implementation section for these procedures is needed.

Of course, the newly created unit *Unit1* must be added in the '*uses*' section of *SpiderUnit.pas* file.

Now, we are ready to create our two primitives, *BmiWO* and *BmiW*:

```
{&P}
procedure BmiWO;
 begin
  BmiWO;
 end;
{&P}
procedure BmiW(kg,cm:Integer);
 begin
  BmiW(kg,cm);
end;
```

These primitives can be used anywhere in the CN as any other primitive.

*A. Python in CNP*

Python is a language very different from Pascal and achieving interoperability with it is not easy. In our approach explained below we use a feature of Free Pascal which allows external programs to be run inside *Lazarus*. This is achieved by importing a *Lazarus* unit called '*process*' which allows a string to be run as if it is run in the program prompt [21, 22]. In our case this string will call the Python interpreter.

We need to start with creating two Python modules: *BmiWO.py* and *BmiW.py* whose codes are given below.

BmiWO.py:

```
kg = int(input('  Please insert your weight in kg '))
cm = int(input('  Please insert your heigth in cm '))
m = (cm / 100)
bmi = ( kg / (m * m) )
print('\n')
if (bmi < 19):
    print('  Your BMI result is :' + str(bmi) + '. You are: Thin ')
elif (bmi < 25):
    print('  Your BMI result is :' + str(bmi) + '. You are: Healthy ')
elif (bmi < 30):
    print('  Your BMI result is :' + str(bmi) + '. You are: Overweight ')
else:
    print('  Your BMI result is :' + str(bmi) + '. You are: Obese ')
```

BmiW.py:

```
import sys
kg = sys.argv[1]
cm = sys.argv[2]
kg = int(kg)
cm = int(cm)
m = (cm / 100)
bmi = ( kg / (m * m) )
print('\n')
if (bmi < 19):
    print('  Your BMI result is :' + str(bmi) + '. You are: Thin ')
elif (bmi < 25):
    print('  Your BMI result is :' + str(bmi) + '. You are: Healthy ')
elif (bmi < 30):
    print('  Your BMI result is :' + str(bmi) + '. You are: Overweight ')
else:
    print('  Your BMI result is :' + str(bmi) + '. You are: Obese ')
print('\n ')
```

We are now ready to create in *SpiderUnit.pas* the corresponding two primitives *BmiWO* and *BmiW*:

```
var
  r: TProcess;

implementation
{&P}
 procedure BmiWO;
 begin
 r:= TProcess.Create(nil);
  r.Options:= r.Options + [poWaitOnExit];
  r.CommandLine:='python ./BmiWO.py';
  r.Execute;
  r.Free;
 end;
{&P}
 procedure BmiW(kg,cm:Integer);
 begin
  r:= TProcess.Create(nil);
  r.Options:= r.Options + [poWaitOnExit];
  r.CommandLine:='python ./BmiW.py';
  r.Execute;
  r.Free;
 end;
```

Calling a process from Free Pascal is implemented as follows. First, a null process is created. *[poWaitOnExit]* is added to the options for synchronization between the process and the execution of the Pascal (CNP) program – the effect of this option is that the Pascal program execution is hold until the process terminates [23]. The corresponding Python module is included in the command line. After the execution of the process it is freed.

It should be noted that the execution of the Python primitive is rather slow as the process involves invoking the Python interpreter.

*B. Java in CNP*

Our approach to interoperability with Java is similar to the usage of Python in CNP. We create a Java program and call it with *TProcess* from *Lazarus* [21]. We wrap this process within primitives so that it can be used in CNP.

The approach is illustrated with the BMI example below.

BmiWO.java:

```
Scanner sc = new Scanner(System.in);
```

```
System.out.println("Please insert your weight in kg");
int kg = sc.nextInt();
System.out.println("Please insert your weight in cm");
int cm = sc.nextInt();
float m = cm;
m = m / 100;
float bmi = ( kg / ( m * m ) );
System.out.println("");
if(bmi < 19){
   System.out.println(" Thin ");
}else if(bmi < 25){
   System.out.println(" Healthy ");
}else if(bmi < 30){
   System.out.println(" Overweight ");
}else{
   System.out.println(" Obese ");
}
   System.out.println("");
```

BmiW.java:

```
String kg = args[0];
String cm = args[1];
float m = Integer.parseInt(cm);
m = m / 100;
float bmi = ((Integer.parseInt(kg)) / ( m * m ));
System.out.println("");
if(bmi < 19){
    System.out.println(" Thin ");
}else if(bmi < 25){
    System.out.println(" Healthy ");
}else if(bmi < 30){
    System.out.println(" Overweight ");
}else{
    System.out.println(" Obese ");
}
System.out.println("");
```

The extra step is the compilation of the .java files and building the .jar files. Then these executable files are wrapped within the primitives. The synchronization of the external process is done using the method explained for the case of Python primitives. An excerpt from *SpiderUnit.pas* follows where the two primitives *BimWO* and *BimW* are defined.

```
var
  r: TProcess;

implementation
 {&P}
 procedure BmiWO;
 begin
   r:= TProcess.Create(nil);
   r.Options:= r.Options + [poWaitOnExit];
   r.CommandLine:='java -jar ./BmiWO.jar';
   r.Execute;
   r.Free;
 end;

 {&P}
 procedure BmiW(kg,cm:Integer);
 begin
   r:= TProcess.Create(nil);
   r.Options:= r.Options + [poWaitOnExit];
   r.CommandLine:='java -jar ./BmiW.jar ' + IntToStr(kg) + ' ' + IntToStr(cm);
   r.Execute;
   r.Free;
end;
```

## V. USING DIFFERENT PROGRAMMING LANGUAGES IN A SINGLE CNP APPLICATION

All the primitives in the CNP application examples in the previous section were written in the same programming language which can be different from Pascal. As a matter of fact, the examples illustrated the usage of C, Python, or Java.

We show below that it is possible to use multiple languages within the same CNP application. We present below new versions of the BMI and Monkey-and-Banana problems.

### A. BMI application with primitives in three different languages

Another BMI calculator is shown below. Its CN is illustrated in Fig, 9.

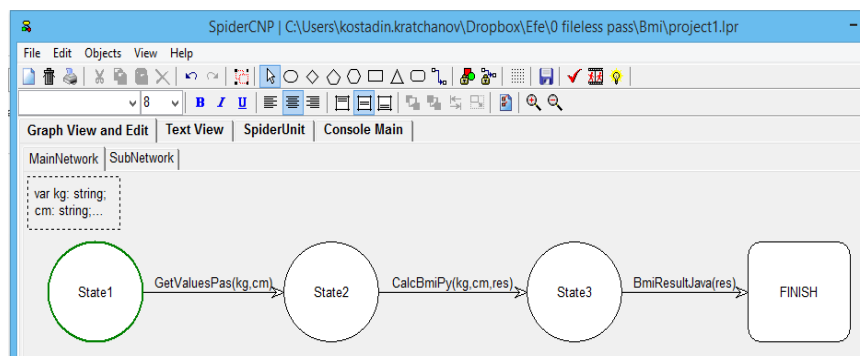

Figure 9. New BMI calculator - CN

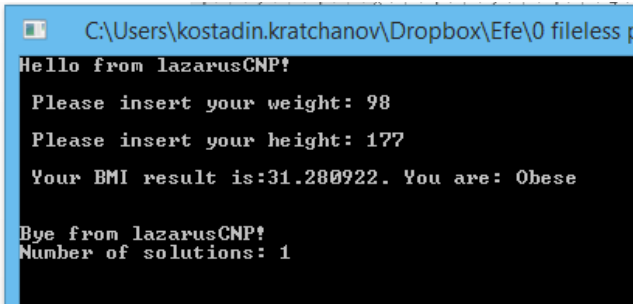

Figure 10. New BMI calculator - Output

Here, primitive *GetValuesPas* is written in Pascal, primitive *CalcBmiPy* – in Python, and primitive *BmiResultJava* – in Java. The file *SpiderUnit.pas* is the following:

*unit Spider Unit;*

```
interface
uses
  SysUtils, Classes, process, Syst;
  function SpiderSolutions : integer;   { for CNP execution ! }
var
  r: TProcess;
  s: TStringList;
implementation
{&P}
 procedure GetValuesPas(var kg,cm:string);
 begin
  Write(' Please insert your weight: ');
  ReadLn(kg);
  WriteLn();
  Write(' Please insert your height: ');
  ReadLn(cm);
  WriteLn();
 end;
 {&P}
 procedure CalcBmiPy(kg,cm: string; var res:string);
 begin
  r:= TProcess.Create(nil);
  r.Options:= r.Options + [poWaitOnExit,poUsePipes];
  r.CommandLine:='python ./BmiCalc.py ' + kg +' '+ cm;
  r.Execute;
  s := TStringList.Create;
  s.LoadFromStream(r.Output);
  r.Free;
  res := s[0];
 end;
{&P}
procedure BmiResultJava(res:string);
begin
   r:= TProcess.Create(nil);
   r.Options:= r.Options + [poWaitOnExit];
   r.CommandLine:='java -jar ./BmiRes.jar ' + res;
   r.Execute;
   r.Free;
end;
```

```
(*&N
*)
function SpiderSolutions : integer;   { for CNP execution ! }
 begin
  SpiderSolutions :=-123
 end;
end.
```

Data between primitives written in different languages is passed using the feature *TProcess* in Free Pascal [21] with option *poUsePipes* which allows the usage of data from the output stream. Data between primitives is passed as strings imitating the system I/O. Option *poWaitOnExit* makes the program wait until the current process terminates which is vital for synchronization between the primitives. Module *BmiCalc.py* is shown in Fig. 11, and file *BmiRes.java* – in Fig. 12.

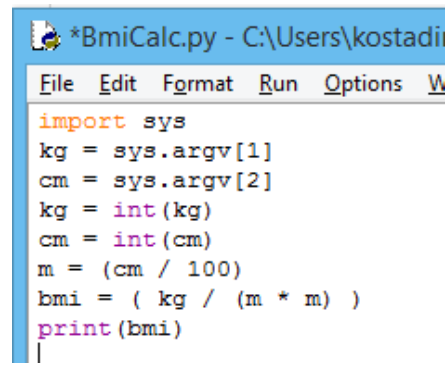

Figure 11. Module BmiCalc.py

```
package bmires;
public class BmiRes {
    public static void main(String[] args) {
        // TODO code application logic here
        float bmi = Float.parseFloat(args[0]);
        if(bmi < 19){
            System.out.println(" Your BMI result is:" + bmi + ". You are: Thin ");
        }else if(bmi < 25){
            System.out.println(" Your BMI result is:" + bmi + ". You are: Healthy ");
        }else if(bmi < 30){
            System.out.println(" Your BMI result is:" + bmi + ". You are: Overweight ");
        }else{
            System.out.println(" Your BMI result is:" + bmi + ". You are: Obese ");
        }
    }
}
```

Figure 12. File BmiRes.java

### B. The Monkey-and-banana application with primitives in four different languages

The 'classical', pure Pascal version of Monkey-and-Banana was discussed in Section II.B above. This new, multi-language version behaves exactly as the 'classical', single language Pascal version and has the same CN! Simply for convenience of the exposition we have changed the names of the primitives so that their new names also indicate the language in which they are written. The CN is specified in Fig. 13 and Fig. 14.

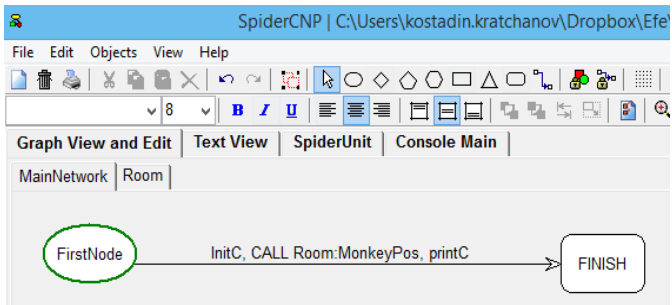

Figure 13. New MB - main subnet

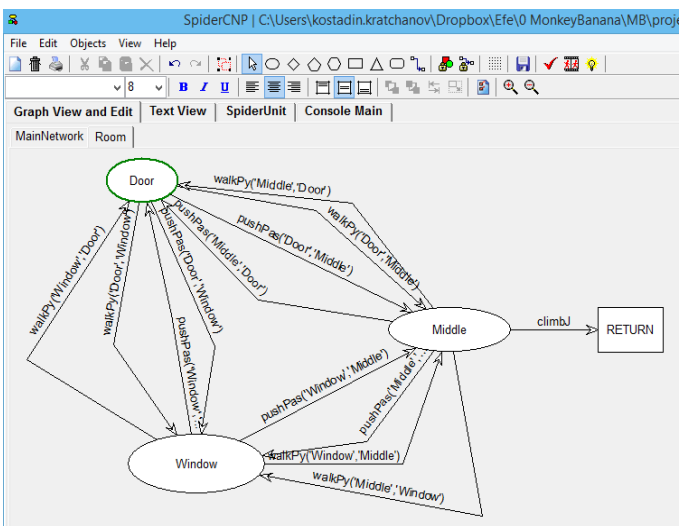

Figure 14. New MB – Room subnet

The language in which the five primitives used are written is as follows:

- C for *InitC* and *PrintC* (because their functions are simple and C is an appropriate choice)
- Python for *WalkPy* (because it is easy to pass strings directly between the main system and Python)
- Pascal for *PushPas* (this is the most complex action and it is convenient to implement it in the native language of the system)
- Java for *ClimbJ* (this is a relatively simple primitive with a certain test)

Although our explanation above seems simple and attractive, its actual implementation is not so easy and straightforward as a developer would desire. The problem lies in the exchange of data between the primitives and the need of the CNP system to keep information about the CN, the current position of the execution and the possible movement backwards. As we stated at the very beginning, this report is only a first attempt into realizing interoperability in CNP. It should be also emphasized that the languages we use do not share a common intermediate language and data system that would support a proper deeper interoperability. As a matter of fact, one of our languages (Python) is even an interpreted language.

Our approach for using primitives written in a language different from Pascal uses the guidelines explained in Section IV. The complete code of the discussed Monkey-and-Banana application is presented in [24]. Below we only briefly discuss the implementation of each of the primitives and the communication between them and between the primitives and the system.

Primitive *InitC* uses a C function *initial*. It performs a dialogue with the user about the positions of the monkey and the banana and stores them a text file *positions.txt*. Then the primitive uses some Pascal code to make the two positions known within the system in order to be used later when other primitives are invoked.

*WalkPy* uses two Python modules, *walkFW* and *walkBW*, which imitate forward, respectively backwards execution along the arrow.

*PushPas* is written in Pascal only. As any action primitive in CNP it has forward and backward parts. It also uses the file *positions.txt* and registers the movement in *steps.txt*.

As known, primitives in CNP are either test (condition) primitives, or action primitives [7, 11]. Action primitives have both a forward execution part and a backwards execution part while the test primitives do not need a backward execution part. It is also possible to have more complicated primitives of a combined nature. For our example, it would be logical on the arrow from state *Middle* to *RETURN* to use a simple test primitive *ComparePositions* which checks if the positions of the monkey and the box coincide, followed by an action primitive *Climb* which registers the climbing step in the *steps.txt* file.

We have preferred to use a combined primitive which includes both functions. Our primitive *ClimbJ* uses the Java packages *ClimbForw.java* and *ClimbBack.java*. The former amends correspondingly the *steps.txt* file by appending the climbing step. It also checks the positions of the monkey and the banana. Correspondingly, the Java code produces outputs 1 or 0. The Java code in *ClmbBack.java* reverses the climbing action by deleting the climbing step from *steps.txt*. Using the *TProcess* feature, the primitive *ClimbJ* calls *ClimbForw.jar* or *ClimbBack.jar* as necessary. It also sets the value of the system variable *Failure* to *true* or *false* depending on the output produced by *ClimbForw.jar* where the test is performed. As a result, if the position of the monkey is not identical to the position of the box the *true* value of the system flag *Failure* will trigger backward movement.

For clarity, the codes of primitive *ClimbJ* and Java package c*limbforw.java* are included below. By the way, the example also illustrates both methods for communication between different languages – using external files (*steps.txt*), and using the system output stream.

CNP primitive *ClimbJ*:

```
{&P}
procedure ClimbJ; //java
 begin
              // runs the java code and reads
the output. then uses the data for decision.
        if forw then begin
              runn := TProcess.Create(nil);
              runn.Options:= runn.Options +
[poWaitOnExit,poUsePipes];
              runn.CommandLine:='java -jar
./ClimbForw.jar ';
              runn.Execute;
                sl := TStringList.Create;
sl.LoadFromStream(runn.Output);
              runn.Free;
              results:= sl[0];
              // decisions happens here
              if results='1' then Failure:=
false else Failure:= true;
        end //forwards
else begin //backwards
              runn := TProcess.Create(nil);
              runn.Options:= runn.Options +
[poWaitOnExit];
              runn.CommandLine:='java -jar
./ClimbBack.jar ';
              runn.Execute;
              runn.Free;
        end //backwards
 end; {Climb}
```

Java package *climbforw.java*:

```
package climbforw;
import java.io.BufferedReader;
import java.io.BufferedWriter;
import java.io.FileNotFoundException;
import java.io.FileReader;
import java.io.FileWriter;
import java.io.IOException;
public class ClimbForw {
    public static void main(String[] args) throws
FileNotFoundException, IOException {
        // reading the positions
        FileReader reader = new
FileReader("./positions.txt");
        BufferedReader br = new
BufferedReader(reader);
        String positions = br.readLine();
        String[] array = positions.split(" ");
        String monkeyPos = array[0];
        String boxPos = array[1];
        FileWriter fw = new
FileWriter("./steps.txt", true);
        BufferedWriter bw = new
BufferedWriter(fw);
        bw.newLine();
        bw.append("Climb");
        bw.newLine();
        bw.append(".");
        bw.close();
        if(monkeyPos.equals(boxPos)){
            System.out.println("1");
        }else {System.out.println("0");    }  }  }
```

Primitive *printC* simply calls the C function *printer* which prints out the steps of the solution from the file *steps.txt*. Both C functions *initial* and *printer* are physically situated in the file *newinit.c*.

## VI. CALLING EXTERNAL PROGRAMS IN CNP

At the end of this exposition we would like to mention that it is possible, from a CN program, to call an application which is completely external to the CNP project.

In the example below, from a CNP application we start the standard Windows calculator. This is done by invoking the executable program *calc.exe*. Of course, in a similar manner, we could call any executable code. The CN of the particular CNP project is given in Fig. 15, and an exemplary output – in Fig. 16.

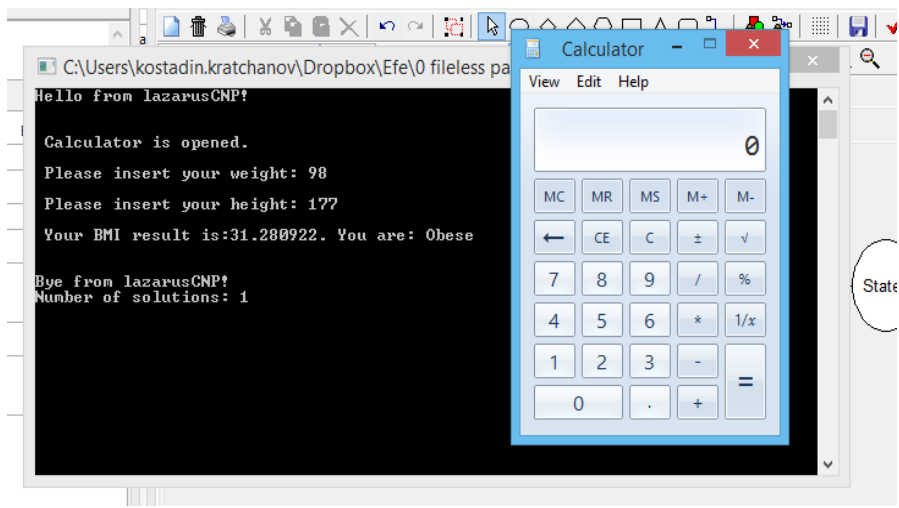

Figure 15. Calculator + BMI – CN

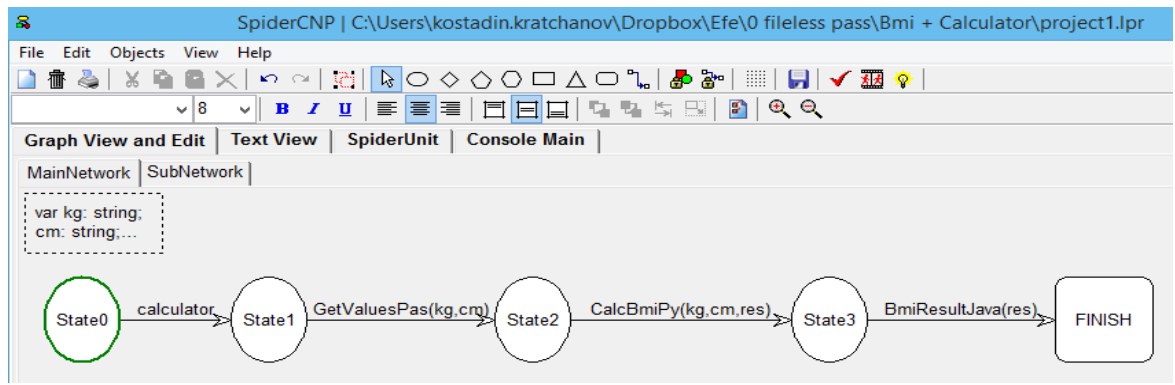

Figure 16. Calculator + BMI - output

The application will start the standard calculator, and then start also the BMI calculator discussed earlier. The code of the *calculator* primitive is shown below:

```
{&P}
procedure calculator;
begin
   r:= TProcess.Create(nil);
   r.CommandLine:='calc.exe';
   r.Execute;
   WriteLn();
   WriteLn(' Calculator is opened. ');
   WriteLn();
   r.Free;
end;
```

As in previous examples, the *TProcess* feature [21] of *Lazarus* is used. This time, however, we have not added the option *[poWaitOnExit]*. Therefore, the Windows calculator and the BMI calculator are not synchronized and can be used simultaneously in parallel. This can be seen in Fig. 10b.

## VII. CONCLUSIONS AND FURTHER RESEARCH

The basics of language interoperability for CNP were developed and presented. The great importance of interoperability for CNP was emphasized and it was shown that the principles of interoperability fit very naturally into the whole idea of CNP. The usage of some most popular non-native languages (C, Java, and Python) for coding the CNP primitives was discussed in detail.

The languages addressed do not share any common virtual machine or data representation. In addition, Python is even an interpreted language. It was shown that still their simultaneous usage is possible. However, the interoperability achieved is not deep enough, it is not seedless and strait forward. Naturally, the execution speed of the applications also suffered. CNP is, first of all, an approach for easy and fast application development, and the interoperability demonstrated contributes substantially in this direction.

A mechanism for writing primitives in any of these languages was described in detail. A method for synchronizing the processes was shown. A few different methods for exchanging data between multiple-language primitives were explained.

This research is only a starting step in developing effective methods for interoperability in CNP, as well as studying the related difficulties and restrictions.

Obviously, the work can be extended to additional programming languages. The first candidates in such a development could be C#, C++, Visual Basic .NET [18] and Kotlin [25].

An important future step would be the construction of CNP IDEs based on languages such as C# and Java where a deeper interoperability could be sought. Another direction for future efforts could also be extending and integrating existing popular advanced IDEs with means for CNP. One more promising area for future efforts is the development of online and cloud-based multi-language CNP environments extending the work reported in [7], as well as light-weight and stand-alone CNP development tools and applications.

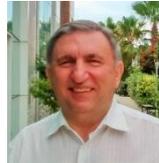

**Kostadin Kratchanov** was born in Kazanlak, Bulgaria. He graduated from the National Technical University of Ukraine - Kyiv Polytechnic Institute in 1974 with a joint BEng/MSc degree in Computer Engineering. He completed a postgraduate qualification in Applied Mathematics and Computer Science in Technical University of Sofia, Bulgaria where he also obtained his PhD degree in Theoretical Computer Science (Reduction of partial fuzzy automata in a category).

He has taught in Computer science and Computer engineering departments at the University of Ruse, Bulgaria, the Technical University of Sofia Plovdiv Branch, Bulgaria, Brunel University, UK, European University of Lefke, Northern Cyprus, University of Bahrain, Grande Prairie Regional College, Canada, and Mount Royal University, Canada. Kostadin is just about to complete his ten-year service as an associate professor in the Software Engineering Department of Yaşar University, Izmir, Turkey. His address for correspondence will be P.O. Box 15, Sofia 1766, Bulgaria.

During his academic career Dr Kratchanov has taken various administrative positions such as head of department, dean, coordinator of EU TEMPUS project for establishing a model information technology faculty. His research interests lie in the areas of theoretical computer science, automata, fuzzy systems, programming languages, programming paradigms and environments.

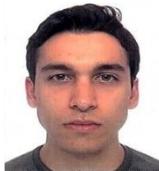

**Efe Ergün** was born in Izmir, Turkey. He graduated with a BSc degree in Software Engineering from Yaşar University in 2018, after spending a period as an Erasmus student at AGH University of Science and Technology, Krakow, Poland. The current paper is based on Efe's senior project which he completed under the supervision of Dr Kratchanov.

Mr Ergün worked as a Software engineerıng intern in Doub.co. He was a radio show host of the Yaşar University web radio. He is currently planning an academic career in artificial intelligence and developing his skills in this field.